\documentclass[aps,10pt,twocolumn,showpacs,nofootinbib]{revtex4-1}

\usepackage{amsmath}
\usepackage{amsfonts}
\usepackage{amssymb}
\usepackage{graphicx}
\usepackage{subfigure}
\usepackage[pdfborder={0 0 0}]{hyperref}  
\usepackage{color}
\usepackage[latin1]{inputenc}
\usepackage{float}

\begin{document}
\title{Quantum entanglement in trimer spin-$1/2$ Heisenberg chains with antiferromagnetic coupling}
 
\begin{abstract}
The quantum entanglement measure is determined, for the first time, for antiferromagnetic trimer spin-$1/2$ Heisenberg chains. The physical quantity proposed here to measure the entanglement is the distance between states by adopting the Hilbert-Schmidt norm. The method is applied to the new magnetic Cu(II) trimer system, $\mathrm{2b.3CuCl_2.2H_2O}$, and to the trinuclear Cu(II) halide salt, $\mathrm{(3MAP)_2Cu_2Cl_8}$.
The decoherence temperature, above which the entanglement is suppressed, is determined for the both systems. A correlation among their decoherence temperatures and their respective exchange coupling constants is established.
\end{abstract}

\author{O.M. Del Cima}
\email{oswaldo.delcima@ufv.br}
\author{D.H.T. Franco}
\email{daniel.franco@ufv.br}
\author{S.L.L. da Silva}
\email{saulo.silva@ufv.br}

\affiliation{Universidade Federal de Vi\c cosa (UFV),\\
Departamento de F\'\i sica - Campus Universit\'ario,\\
Avenida Peter Henry Rolfs s/n - 36570-900 -
Vi\c cosa - MG - Brazil.}


\maketitle

\section{Introduction}
Since the seminal works by Einstein, Podolski and Rosen \cite{ref1.1}, and by Schr\"odinger \cite{schrodinger}, quantum entanglement 
has became one of the most amazing and pursued phenomenon in quantum mechanics, mainly, after 
the experiment realized by Aspect, Grangier and Roger \cite{ref3}. During the last decade, quantum entanglement has been extremely important in processing and transmission of information \cite{ref1.2, ref1.3}, in understanding the quantum phase transitions \cite{ref1.4, ref1.5}, as well as in applications in quantum biology \cite{ref1.6}. 

Very recently, an experimental realization of long-distance entanglement in antiferromagnetic spin chains has been performed \cite{long-distance}. However, in spite of being an old and deeply studied subject, from theoretical and experimental point of view, there still has a lack of a measure of entanglement -- there is no quantitative method -- a physical quantity which measures the degree of quantum entanglement. Qualitative criteria are available, where in some special cases, detect whether or not the entanglement is presented, without foreseen how much a system is entangled. Since entanglement is a quantum mechanical phenomenon, it should not to be revealed at scales larger than atomic or in high temperatures, this is known as quantum decoherence. However, as it has been presented here, there are solids exhibiting quantum entanglement even at finite temperatures -- the trimer spin-$1/2$ Heisenberg chain compounds with antiferromagnetic exchange coupling, $\mathrm{2b.3CuCl_2.2H_2O}$ \cite{ref7} and $\mathrm{(3MAP)_2Cu_2Cl_8}$ \cite{ref9}. 
As a physical quantity to measure  
quantum entanglement, use has been made of the distance between states \cite{ref1.7, ref1.8} by using the Hilbert-Schmidt norm \cite{ref5}.

The outline of this paper is as follows. In Section \ref{sec:Calcu}, the method adopted to compute the quantum entanglement is introduced and its measure, the distance among states, is established for any spin-$1/2$ antiferromagnetic trimer. The quantum entanglement, for the compounds  $\mathrm{2b.3CuCl_2.2H_2O}$ and $\mathrm{(3MAP)_2Cu_2Cl_8}$, and their critical temperatures, above which the entanglement vanishes, are computed and presented in Section \ref{sec:Comp}. Finally, Section \ref{sec:Conc} is left for conclusions.

\section{Measuring the quantum entanglement}
A antiferromagnetic trimer spin-$1/2$ Heisenberg chain can be represented by the Hamiltonian:
\begin{equation}
H_{\rm AF} = -J \sum_i \vec{S_i} \cdot \vec{S}_{i+1}~, \label{H}
\end{equation}
where $J<0$ is the exchange coupling constant and $S_i$ is the spin operator for the $i$th spin (on site $i$). The Hamiltonian (\ref{H}) commutes with the ${\rm z}$-component of the total spin ($S^{\rm z}$), $[H_{\rm AF},S^{\rm z}]=0$. Therefore, the reduced density matrix can be written as follows \cite{ref1}:
\begin{equation}
\rho_{i,i+1}=
\begin{pmatrix}
v^{+} & 0 & 0 & 0 \\
0 & w & z & 0 \\
0 & z^* & x & 0 \\
0 & 0 & 0 & v^{-}
\end{pmatrix}~, \label{dmatrix}
\end{equation}
such that the indices $i$ and $i+1$ refer to the site $i$, and its nearest neighbour site, $i+1$, respectively. Since there is no coherent superposition of states and the exchange coupling is antiferromagnetic, the density matrix (\ref{dmatrix}) is rewritten as
\begin{equation}
\rho_{i,i+1}=
\begin{pmatrix}
v & 0 & 0 & 0 \\
0 & w & z & 0 \\
0 & z^* & w & 0 \\
0 & 0 & 0 & v
\end{pmatrix}~. \label{dmatrix1}
\end{equation}
The elements of the reduced density matrix (\ref{dmatrix1}) can be related with the correlation functions (per site) \cite{ref2}, and they read 
\begin{eqnarray}\label{eq3}
&& v=\frac{1}{4}+\langle S_i^{\rm z}S_{i+1}^{\rm z}\rangle~,\\
&& z=\langle S_i^{\rm x}S_{i+1}^{\rm x}\rangle + \langle S_i^{\rm y}S_{i+1}^{\rm y}\rangle + 
i\langle S_i^{\rm x}S_{i+1}^{\rm y}\rangle - i\langle S_i^{\rm y}S_{i+1}^{\rm x}\rangle
\end{eqnarray}
and since ${\rm Tr}(\rho_{i,i+1})=1$,
\begin{equation}
w=\frac{1}{2}(1-2v)~.
\end{equation}
However, the system is isotropic when in the absence of external magnetic field, so
\begin{eqnarray}\label{eq4}
&& v=\frac{1}{4}+\langle S_i S_{i+1}\rangle~, ~~z=2\langle S_iS_{i+1} \rangle~, \nonumber\\
&& w=\frac{1}{4}-\langle S_i S_{i+1}\rangle~. \label{vwz}
\end{eqnarray}

Since the Hilbert space dimension is of a $2\otimes2$-system, $\rho_{i,i+1}=(\rho^{A}\otimes\rho^{B})_{i,i+1}$, the Peres-Horodecki criterion \cite{ref4} shall be used. The eigenvalues, $\lambda_k=(\lambda_1,\lambda_2,\lambda_3,\lambda_4)$, of the partial transposition of $\rho_{i,i+1}$ (\ref{dmatrix1}), $\rho_{i,i+1}^{{\rm T}_B}$, are
\begin{equation}
\lambda_k=(w, w, v+|z|, v-|z|)~. \label{eigenv}
\end{equation}
However, from (\ref{vwz}) and (\ref{eigenv}), together to the fact that
\begin{equation}
\langle S_iS_{i+1} \rangle=\frac{1}{2}\langle (S_i+S_{i+1})^2 \rangle - \frac{3}{4} \label{SiSi+1}
\end{equation} 
for $T=0\mathrm{K}$, then,
\begin{equation}
-\frac{3}{4}\leq\langle S_iS_{i+1} \rangle\leq-\frac{1}{4}~, \label{<SiSi+1<}
\end{equation}
which yields that $\lambda_1=\lambda_2>0$ and $\lambda_3>0$. The Peres-Horodecki criterion states that, a 
system is entangled if, at least, one of the eigenvalues $\lambda_k$ (\ref{eigenv}) is negative, otherwise, it is separable. 
It remains one eigenvalue to be analysed, $\lambda_4=v-|z|$. Taking into account (\ref{vwz}), (\ref{eigenv}) and (\ref{<SiSi+1<}), it stems that $v < |z|$, which results $\lambda_4<0$, therefore, as a consequence from the Peres-Horodecki criterion \cite{ref4}, the system is fully entangled for $T=0\mathrm{K}$. 

It shall be stressed that, from now on, by knowing the separate and the entangled states, it can calculated how much a state is entangled. 
The distance between states \cite{ref1.7, ref1.8} has been adopted to measure the entanglement through the Hilbert-Schmidt norm \cite{ref5}. 
Therefore, the entanglement measure is given by
\begin{equation}
\mathcal{E}(\rho)=\mathrm{min}\,D(\rho_s \vert \vert \rho_e)~, \label{measure}
\end{equation}
where $D$ is the distance between separable and entangled states and, $\rho_s$ and $\rho_e$ are their reduced density matrices, 
respectively. In other words, the entanglement is measured as the closest distance between a given entangled 
state and the set of separable states. The reduced density matrices,  $\rho_e$ e $\rho_s$, associated to entangled and 
separable states, respectively, read
\begin{equation}
\rho_e=
\begin{pmatrix}
v_{_<} & 0 & 0 & 0 \\
0 & w_{_<} & z & 0 \\
0 & z^* & w_{_<} & 0 \\
0 & 0 & 0 & v_{_<}
\end{pmatrix}\label{matrixe}
\end{equation} 
and
\begin{equation}
\rho_s=
\begin{pmatrix}
v_{_\geqslant} & 0 & 0 & 0 \\
0 & w_{_\geqslant} & z & 0 \\
0 & z^* & w_{_\geqslant} & 0 \\
0 & 0 & 0 & v_{_\geqslant}
\end{pmatrix}~,\label{matrixs}
\end{equation}
where the subscripts, $<$ and $\geqslant$, indicate that the condition, $v<|z|$, is satisfied by the former and does not 
by the latter -- which should happen at finite temperatures, $T>0\mathrm{K}$. 

The measure of entanglement $\mathcal{E}(\rho)$ (\ref{measure}), by assuming the Hilbert-Schmidt norm \cite{ref5}, 
is now written as 
\begin{eqnarray}\label{eq2}
\mathcal{E}(\rho)&=&\mathcal{E}_0\,\mathrm{min}\sqrt{{\rm Tr}[(\rho_s - \rho_e)^2]} \nonumber\\
&=&2\,\mathcal{E}_0\,\mathrm{min}\vert v_{_\geqslant} - v_{_<} \vert~,
\end{eqnarray}
where $\mathcal{E}_0$ is a normalization constant so that the entanglement measure (\ref{eq2}) satisfies:
\begin{equation}
0\leqslant\mathcal{E}(\rho)\leqslant 1~. \label{0measure1}
\end{equation}
Furhermore, $\mathcal{E}(\rho)$ (\ref{eq2}) reaches its minimum for $v_{_\geqslant}=\vert z \vert$, 
giving rise to
\begin{equation}\label{eq5}
\mathcal{E}(\rho)=\mathcal{E}_0\,\mathrm{max}\left[0,2(\vert z \vert - v_<)\right]~,
\end{equation}
where, in this case, $\mathcal{E}_0={1}/{4}$. Then, by substituting (\ref{vwz}) in (\ref{eq5}), it follows that
\begin{eqnarray}\label{eq6}
&&\mathcal{E}(\rho)= \nonumber\\ 
&&2\,\mathcal{E}_0\,\mathrm{max}\left[0,\left(2\vert \langle S_iS_{i+1}\rangle \vert - 
\frac{1}{4} - \langle S_iS_{i+1}\rangle\right) \right]~.
\end{eqnarray} 

Due to the fact that the Hamiltonian $H_{\rm AF}$ (\ref{H}) commutes with the spin component along the $z$
direction $S^{\rm z}$, the magnetic susceptibility \cite{ref6} along a given direction $\alpha$ ($\chi^\alpha(T)$) 
can be written as
\begin{equation}\label{eq7}
\chi^\alpha(T)=\frac{(g\mu_{\rm{B}})^2}{k_{\rm{B}}T}\left(\sum_{i,j=1}^{N}\langle S_i^\alpha S_j^\alpha\rangle - 
\left\langle \sum_{i=1}^{N}S_i^\alpha \right\rangle^2 \right)~,
\end{equation}
with
\begin{equation}\label{eq8}
\overline{\chi}(T)=\frac{\chi^{\rm x} +\chi^{\rm y} +\chi^{\rm z}}{3}~,
\end{equation}
being the average of the magnetic susceptibility measured along the three orthogonal axis. It should be noticed that for optical lattices \cite{optical-lattices}, the variances in (\ref{eq7}) can be directly measured without to stand in need of the magnetic susceptibility. 

In the following Section, the quantum entanglement measure $\mathcal{E}(\rho)$ have been computed for the trimer compounds $\mathrm{2b.3CuCl_2.2H_2O}$ \cite{ref7} and $\mathrm{(3MAP)_2Cu_2Cl_8}$ \cite{ref9}, and their respective entanglement critical temperatures determined.   

\label{sec:Calcu}

\section{Quantum entanglement measure and antiferromagnetic trimer systems}
\begin{figure}[ht!]
 \centering
 \includegraphics[scale=0.70]{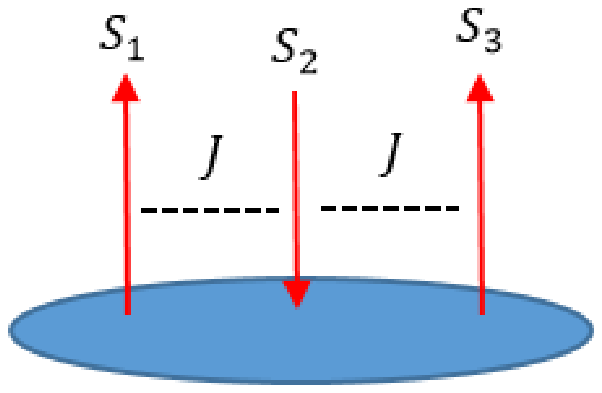}
 \caption{An antiferromagnetic trimer system with exchange coupling constant $J$.}
 \label{f:1}
\end{figure}

Any antiferromagnetic trimer system, with exchange coupling constant $J$, as represented in Figure \ref{f:1}, can be modelled as proposed here. Despite there is also an exchange interaction among the trimers 
($J_{\rm int}$), such interaction is very small if compared with the intra-trimer exchange interaction 
($J$), $J_{\rm int}\ll J$, so it has not been taken into account.
A general approach to calculate the quantum entanglement measure, $\mathcal{E}(\rho)$, has to be established before its further application for the compounds $\mathrm{2b.3CuCl_2.2H_2O}$ \cite{ref7} and $\mathrm{(3MAP)_2Cu_2Cl_8}$ \cite{ref9}.  
Bearing in mind (\ref{eq7}) and (\ref{eq8}), it straightly follows that
\begin{equation}\label{eq9}
\chi^{\rm z}(T)=\frac{2(g\mu_{\rm{B}})^2}{k_{\rm{B}}T}(2\langle S_iS_{i+1}\rangle+1)~,
\end{equation}
then
\begin{equation}\label{eq10}
\overline{\chi}(T)=\frac{2(g\mu_{\rm{B}})^2}{3k_{\rm{B}}T}(2\langle S_iS_{i+1}\rangle+1)~.
\end{equation}
Therefore, from (\ref{eq10}) and (\ref{eq6}), the quantum entanglement measure $\mathcal{E}(\rho)$, reads 
\begin{align}\label{eq11}
\mathcal{E}(\rho)=\mathcal{E}_0\frac{k_{\rm{B}}T}{(g\mu_{\rm{B}})^2}\,\mathrm{max}
&\left[0,\left(2\left| \frac{3}{2}\overline{\chi}(T)
-\frac{(g\mu_{\rm{B}})^2}{k_{\rm{B}}T} \right| \right. \right. \nonumber \\[3mm] 
&+\left. \left.\frac{(g\mu_{\rm{B}})^2}{2k_{\rm{B}}T}-\frac{3}{2}\overline{\chi}(T) \right) \right]~.
\end{align}
The mean susceptibility value $\overline{\chi}(T)$ for the trimer system, with antiferromagnetic exchange coupling constant $J$, has been calculated by using the Van Vleck equation \cite{ref8}, yielding that
\begin{equation}\label{eq12}
\overline{\chi}(T)=\frac{(g\mu_{\rm{B}})^2}{4k_{\rm{B}}T}
\frac{1+e^{\frac{J}{k_{\rm{B}}T}}+10e^{\frac{3J}{2k_{\rm{B}}T}}}{1+e^{\frac{J}{k_{\rm{B}}T}}+2e^{\frac{3J}{2k_{\rm{B}}T}}}~.
\end{equation}
Finally, by substituting (\ref{eq12}) in (\ref{eq11}), the quantum entanglement measure 
($\mathcal{E}(\rho)$) is attained as a function of the temperature ($T$) and the exchange coupling ($J$):
\begin{eqnarray}
\mathcal{E}(J,T)&=&\mathcal{E}_0\,\mathrm{max}\left[0,\left(2\left|\frac{3}{8}\frac{1+e^{\frac{J}{k_{\rm{B}}T}}+
10e^{\frac{3J}{2k_{\rm{B}}T}}}{1+e^{\frac{J}{k_{\rm{B}}T}}+2e^{\frac{3J}{2k_{\rm{B}}T}}}
-1 \right| \right.\right. \nonumber \\[3mm] 
&+&\frac{1}{2}-\left.\left.\frac{3}{8}\frac{1+e^{\frac{J}{k_{\rm{B}}T}}+10e^{\frac{3J}{2k_{\rm{B}}T}}}
{1+e^{\frac{J}{k_{\rm{B}}T}}+2e^{\frac{3J}{2k_{\rm{B}}T}}} \right) \right]~,\label{measureJT}
\end{eqnarray}
with
\begin{equation}
\lim_{T\rightarrow 0}{\mathcal{E}(J,T)}=\frac{11}{8}\mathcal{E}_0=\frac{11}{32}~.
\end{equation}
Moreover, in which concerns the result above (\ref{measureJT}), it should be stressed that the entanglement critical temperature ($T_{c}$) is defined as the one which vanishes identically the quantum entanglement measure, $\mathcal{E}(J,T)$: 
\begin{equation}
\left.\mathcal{E}(J,T)\right|_{T=T_{c}}\equiv 0~.\label{Tc}
\end{equation}
Thus, for temperatures below the critical (decoherence) one, $T<T_{c}$, the system is entangled, whereas, for temperatures above it, $T\geqslant T_{c}$, the system experiences quantum decoherence, it assumes a separable state. 

\begin{figure}[ht!]
 \centering
 \includegraphics[scale=0.90]{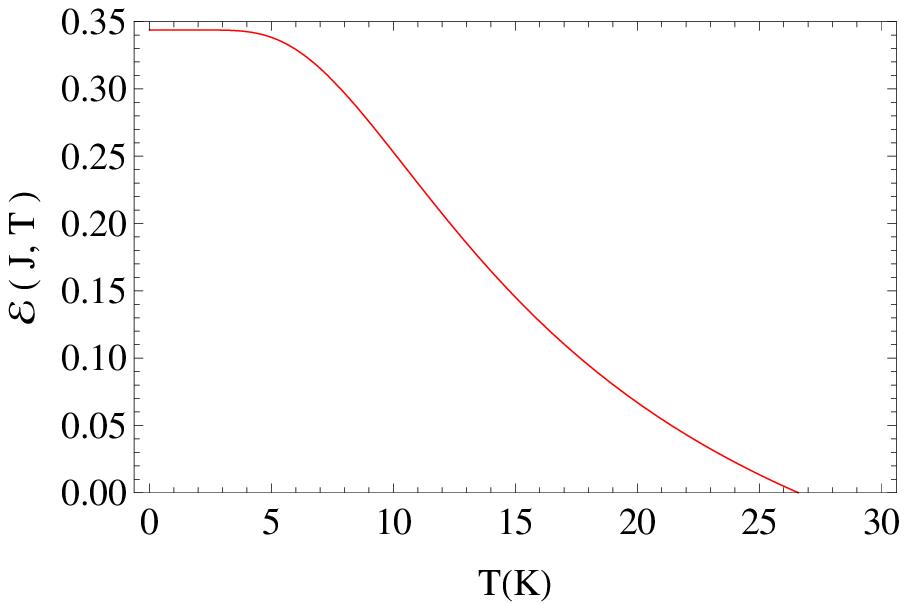}
 \caption{The quantum entanglement measure, $\mathcal{E}(J,T)$, for $\mathrm{2b.3CuCl_2.2H_2O}$ \cite{ref7}, with $J/k_{\rm{B}}=-20,0 \mathrm{K}$, has $T_{c}=26,6\mathrm{K}$.}
 \label{f:3}
\end{figure}

\begin{figure}[ht!]
 \centering
 \includegraphics[scale=0.90]{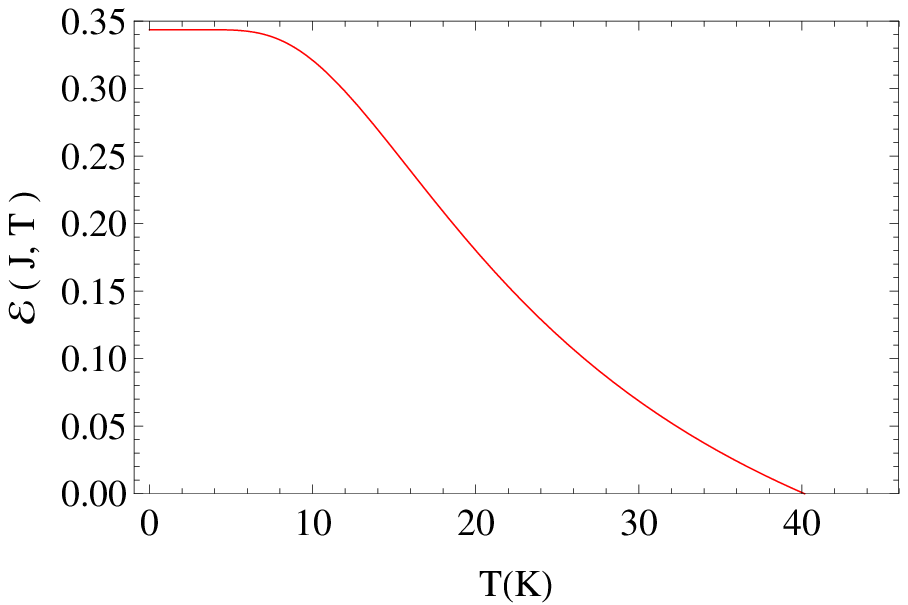}
 \caption{The quantum entanglement measure, $\mathcal{E}(J,T)$, for $\mathrm{(3MAP)_2Cu_2Cl_8}$ \cite{ref9}, with $J/k_{\rm{B}}=-30,2 \mathrm{K}$, has $T_{c}=40,2\mathrm{K}$.}
 \label{f:4}
\end{figure}

The application to any trimer system with antiferromagnetic interactions, in order to determinate 
the degree of entanglement (quantum entanglement measure) and the decoherence temperature (critical temperature) is now straightforward. The first material analysed was the new magnetic Cu(II) trimer system  $\mathrm{2b.3CuCl_2.2H_2O}$\,\footnote{$b=\mathrm{C_5H_{11}NO_2}$ (betaine). For details concerning   
crystal structure and magnetic properties, see \cite{ref7}} \cite{ref7}, 
where $J/k_{\rm{B}}=-20,0\mathrm{K}$. The quantum entanglement measure, $\mathcal{E}(J,T)$ (\ref{measureJT}), 
of $\mathrm{2b.3CuCl_2.2H_2O}$, as a function of temperature is displayed in Figure \ref{f:3}. The critical temperature obtained for $\mathrm{2b.3CuCl_2.2H_2O}$, from solving (\ref{measureJT}) together with (\ref{Tc}), 
is $T_{c}=26,6\mathrm{K}$.
The second material analysed was a trinuclear Cu(II) halide salt 
$\mathrm{(3MAP)_2Cu_2Cl_8}$\,\footnote{3MAP=3-methyl-2aminopyridinium. For details concerning crystal structure and magnetic properties, see \cite{ref9}.} \cite{ref9}, where $J/k_{\rm{B}}=-30,2\mathrm{K}$. The quantum entanglement measure, $\mathcal{E}(J,T)$ (\ref{measureJT}), of $\mathrm{(3MAP)_2Cu_2Cl_8}$, as a function of temperature is presented  in Figure \ref{f:4}, and its critical temperature (decoherence temperature), calculated in the same way as the first compound, is $T_{c}=40,2\mathrm{K}$.

Comparing the results obtained for $\mathrm{2b.3CuCl_2.2H_2O}$ (Figure \ref{f:3}) and 
$\mathrm{(3MAP)_2Cu_2Cl_8}$ (Figure \ref{f:4}), an interesting result stems from, 
the greater the absolute value of the exchange coupling constant ($|J|$), the greater the decoherence (critical) temperature ($T_{c}$). As a matter of fact, it can be directly verified, from (\ref{measureJT}) and (\ref{Tc}), that $T_{c}$ is proportional to $|J|$, $T_{c}\propto |J|$.
\label{sec:Comp}
\section{Conclusions}
The quantum entanglement is measured by the distance between states \cite{ref1.7, ref1.8}, 
through the Hilbert-Schmidt norm \cite{ref5}, and established for any spin-$1/2$ antiferromagnetic trimer. It is analytically computed for the new magnetic Cu(II) trimer system  $\mathrm{2b.3CuCl_2.2H_2O}$ ($J/k_{\rm{B}}=-20,0 \mathrm{K}$) \cite{ref7}, and for the trinuclear Cu(II) halide salt 
$\mathrm{(3MAP)_2Cu_2Cl_8}$ ($J/k_{\rm{B}}=-30,2 \mathrm{K}$) \cite{ref9}, as well as their decoherence (critical) 
temperatures, $T_{c}=26,6\mathrm{K}$ and $T_{c}=40,2\mathrm{K}$, respectively. Furthermore, the correlation 
among the decoherence temperature ($T_{c}$) and the absolute value of the exchange coupling constant ($|J|$), $T_{c}\propto |J|$, is explicitly identified -- the greater $|J|$, the 
greater is $T_{c}$. It should be stressed that the quantum entanglement measure is independent on 
the exchange coupling only at $T=0\mathrm{K}$. Therefore, as a conjecture, the exchange coupling protects the system from decoherence as temperature increases. 
\label{sec:Conc}
\\

\begin{acknowledgments}
  The authors thank Afrânio R. Pereira for valuable comments and encouragement, and G\'eza T\'oth for pointed us out the application of the method proposed here to optical lattices. This work was partially supported by the Brazilian agencies, FAPEMIG and CAPES. O.M.D.C. dedicates this work to his father (Oswaldo Del Cima, {\it in memoriam}), mother (Victoria M. Del Cima, {\it in memoriam}), daughter (Vittoria) and son (Enzo).
\end{acknowledgments}

\bibliographystyle{apsrev4-1}
\bibliography{ensembles}

\end{document}